\documentclass[aps,preprint,showpacs,superscriptaddress]{revtex4}
\usepackage{amsfonts,amsmath,graphicx}

\def \o{\omega}
\def \e{\varepsilon}
\def \be{\begin{equation}}
\def \ee{\end{equation}}

\begin{document}

\title{Beam Propagation in Photonic Crystals}
\author{B. Guizal}
\affiliation{ D\'{e}partement d'optique, Institut FEMTO-ST, UMR\ 6174\\
Universit\'{e} de Franche-Comt\'{e}\\
16, Route de Gray 25030 Besan\c{c}on Cedex \ \ France\\ }

\author{D. Felbacq}
\affiliation{Groupe d'Etude des Semiconducteurs \\
Unit\'{e} Mixte de Recherche du Centre National de la Recherche
Scientifique
 5650\\
Universit\'{e} Montpellier II\\
34095 Montpellier Cedex 5, France\\}

\author{R. Sma\^{a}li }
\affiliation{Laboratoire des technologies de la
micro\'{e}lectronique - 17 rue des Martyrs, 38054 Grenoble, Cedex 9
France.}
\date{\today }

\begin{abstract}
The recent interest in the imaging possibilities of photonic
crystals (superlensing, superprism, optical mirages etc...) call for a detailed analysis of beam propagation inside a
finite periodic structure. In this paper, we give such a theoretical
and numerical analysis of  beam propagation in 1D and 2D photonic
crystals. We show that, contrarily to common knowledge, it is not
always true that the direction of propagation of a beam is given by
the normal to the dispersion curve. We explain this phenomenon in
terms of evanescent waves and we construct a renormalized dispersion
curve that gives the correct direction.
\end{abstract}
\maketitle
\section{Introduction and setting of the problem}

Some beautiful experiments and numerical works have shown that it
was possible to obtain quite unusual behaviors of light propagation
inside meta-materials and photonic crystals (PhCs).
\cite{ozbay,tret,fang,moi2,centeno,benisty,matsu,pendry}.
 In particular, recent ideas by Pendry confirmed by experiments show
that photonic crystals, maybe under the guise of meta-materials,
could prove to be of huge interest in that they could allow to beat
the diffraction limit and to make superlenses. The point of this
work is to give a theoretical insight into beam propagation inside
PhCs and in particular on the computation of the shift of the
transmitted beam (cf. Figure 1).
We consider both 1D and 2D PhCs and we show that, contrarily to what is generally
believed, the direction of propagation is not always directly given
by the normal to the dispersion curves for 2D PhCs. Rather, we define a
renormalized, or effective, dispersion diagram, whose normal gives
the correct direction of propagation. Numerical examples are given
illustrating the various regimes.

We will begin by studying, systematically, 1D structures before
extending the results to 2D structures (seen as stacks of gratings)
through the concept of the so-called two waves
approximation that will be introduced later.
Throughout this work, we use time-harmonic fields, with a
time-dependence of $\exp (-i\o t)$, that are $z$-independent.
The vectorial diffraction problem is reduced to the study
of the two usual cases of polarization: $s$-polarization (electric
field parallel to the grooves of the gratings) or $p$-polarization
(magnetic field parallel to the grooves). The wavenumber is denoted
by $k_{0}=\frac{2\pi }{\lambda }$, where $\lambda $ is the incident
wavelength in vacuum.

The incident field is a Gaussian beam whose $z$ component can be
expressed by
\begin{equation}
u^i \left( x,y\right) =\int A\left( \alpha \right) e^{i\left( \alpha x+
\sqrt{k_0^2-\alpha ^2}y\right) }d\alpha
\end{equation}
where
\[
A\left( \alpha \right) =\frac{w}{2\sqrt{\pi
}}e^{-\frac{w^{2}}{4}\left( \alpha -\alpha _{0}\right) ^{2}},
\]
we denote $u^i$ the electric (magnetic) field in the case of $s\, (p)$
-polarization and $\alpha _0=k_0 \sin \theta _{0}$, where $\theta_{0}$
 is the mean angle of incidence of the beam and $w$ its
waist.

\section{Analysis of the beam propagation}

\subsection{The case of a stratified medium : 1D PhC}

In this section, we derive the value of the shift of the transmitted
beam for the particular case of a stratified medium, i.e. when the
relative permittivity is constant in the horizontal direction: it is
described by a real periodic function (period $h$):
$\e \left( y\right) $. We denote: ${\mathbf r}=(x,y)$.
 We consider $N$ periods of the stratified
medium, which is embedded in vacuum. For an incident plane wave of
wavevector $\mathbf{k}=k\left( \sin \theta ,-\cos \theta , 0\right)
$, we denote $\beta _{0}=k_{0}\cos \theta $ and $\left( r_{N}\left(
k,\theta \right) ,t_{N}\left( k,\theta \right) \right) $ the
reflection and transmission coefficients of the structure.
 the electromagnetic field in the outer regions reads as:
\begin{eqnarray}
u({\mathbf r}) &=&e^{i\mathbf{k.r}}+r_{N}e^{ik_{0}\left(x \sin \theta
+y\cos \theta \right) },\,y\geq 0 \\
u({\mathbf r}) &=&t_{N}e^{ik_{0}\left(x \sin \theta -(y+Nh)\cos
\theta \right) },\,y\leq -Nh
\end{eqnarray}

We denote by $\mathbf{T}$ the transfer
matrix of one period, then it is known
\cite{moi1} that the reflection and transmission coefficients are
related through the relation:
\begin{equation}
\mathbf{T}^{N}\left(
\begin{array}{c}
1+r_{N} \\
i\beta _{0}\left( 1-r_{N}\right)
\end{array}
\right) =t_{N}\left(
\begin{array}{c}
1 \\
i\beta _{0}
\end{array}
\right)  \label{transfert}
\end{equation}
\newline
Let us denote by $\gamma$ and $\gamma^{-1}$ the eigenvalues of $\mathbf{T}$ and by
$\mathbf{v}=\left( \phi_{11},\phi _{21}\right) ,\mathbf{w}=\left( \phi _{12},\phi _{22}\right)$
the associated eigenvectors ($\mathbf{T}\mathbf{v}=\gamma \mathbf{v}\, ,\,
\mathbf{T}\mathbf{w}=\gamma^{-1} \mathbf{w}$).
It is known (see for instance \cite{moi1}) that the band gaps and the conduction bands are given respectively by: $\mathbf{G}=\left\{ \left( k,\theta \right) ,tr\left( \mathbf{T}\right) >2\right\} $,
and : $\mathbf{B}=\left\{ \left( k,\theta \right) ,tr\left(
\mathbf{T}\right) <2\right\} $.
The reflection and transmission coefficients are then given by:
\begin{equation}
r_{N}\left( k,\theta \right) = \frac{\left( \gamma^{2 N}-1\right)
f}{ \gamma^{2 N}-g^{-1}f} \, , \,t_{N}\left( k,\theta \right) =
\frac{ \gamma^{N}\left(1-g^{-1}f\right) }{ \gamma^{2 N}-g^{-1}f}
\end{equation}
\newline
where, denoting $q(x,y)={{\frac{i\beta _{0}y-x}{i\beta _{0}y+x}}}$,
the functions $f$ and $g$ are defined by $g\left( k,\theta \right)
=q\left( \mathbf{v}\right) ,f\left( k,\theta \right) =q\left(
\mathbf{w}\right)$ and $\mathbf{v}$ is chosen such that $|g|\leq1$
in the conduction bands. Remark that in these bands, the inverse of
$f$ is equal to the conjugate of $g$ (see \cite{moiprl,dofrescoex} for details).

A straightforward calculation shows that, for $\left( k,\theta \right) \in
\mathbf{B}$:
\begin{eqnarray}
r_{N}\left( k,\theta \right)  &=&g+\left( g-f\right) \sum_{p=1}^{+\infty
}\gamma ^{2Np}\left| g\right| ^{2p} \\
t_{N}\left( k,\theta \right)  &=&(1-|g|^2)\gamma
^{N}\sum_{p=0}^{+\infty }\gamma ^{2Np}\left| g\right| ^{2p}
\end{eqnarray}
In the conduction bands, the eigenvalue $\gamma$ can be written
under the form: $\gamma=e^{i\beta h}$, where $\beta$ is the
so-called Bloch phase. When the incident field is a beam, we get an
infinite sum of transmitted beams, corresponding to multiple
scattering. Let us concentrate on the first transmitted beam, i.e.
the beam that reads:
\begin{equation}
u_{0}^{t}\left( x,Nh\right) =\int A\left( \alpha \right) \left(
1-\left| g\right| ^{2}\right) e^{i\beta hN}e^{i\alpha x}d\alpha
\end{equation}
whose Fourier transform is:
\begin{equation}
\widehat{u}_{t}\left( \alpha \right) =\sqrt{2\pi
}\widetilde{A}\left( \alpha \right) e^{i\beta Nh}
\end{equation}

where
\begin{equation}
\widetilde{A}\left( \alpha \right) =A\left( \alpha \right) \left( 1-\left|
g\right| ^{2}\right) .
\end{equation}

We denote $G_{i},G_{t},G_{d}$ the points where, respectively, the incident, transmitted and
reflected beams enter or emerge from the PhC. Given the incident field,
the axis are chosen to have $G_i=0$. These points are defined as the
barycenters, or first moments, of the corresponding fields, that is:
\begin{equation}
\begin{array}{lll}
G_{i}&=\frac{\int x \left| u^{i}\left( x,0\right) \right| ^{2}dx}{\int \left| u^{i}\left( x,0\right) \right| ^{2}dx}&=0 \\
G_{d}&=\frac{\int x \left| u^{d}\left( x,0\right) \right| ^{2}dx}{\int \left| u^{d}\left( x,Nh\right) \right| ^{2}dx} \\
G_{t}&=\frac{\int x \left| u^{t}\left( x,nh\right) \right| ^{2}dx}{\int \left| u^{t}\left( x,Nh\right) \right| ^{2}dx}
\end{array}
\end{equation}

Using Parseval-Plancherel identity, we get the angular shift due to the beam propagation  (cf.
fig. 1):
\begin{equation}
\tan \psi =\frac{G_{t}}{Nh}=-\frac{\int \widetilde{A}^{2}\left(
\alpha
\right) \partial_{\alpha}\beta \left( \alpha \right) d\alpha }{\int \widetilde{A}^{2}
\left( \alpha \right) d\alpha }
\end{equation}

A series expansion of $\tan \psi$ can be obtained provided the phase function is analytic with respect to
$\alpha $ in a neighborhood of $\alpha_0$. Indeed, we can then write:
\begin{equation}
\partial_{\alpha}\beta\left( \alpha \right) =
\sum_{m=0}^{+\infty}
\frac{\partial_{\alpha}^{m+1} \beta \left( \alpha _{0}\right)}{m!}\left( \alpha -\alpha _{0}\right) ^{m}
\end{equation}

We obtain after some manipulations:
\begin{equation}
\tan \psi =-\sum_{m}\frac{2^{m}\Gamma \left( m+\frac{1}{2}\right)}{\left( 2m\right) !}
\partial_{\alpha }^{2m+1}\beta \left( \alpha_{0}\right)
\end{equation}
where $\Gamma$ is the Euler Gamma function \cite{abram}.
When $k_{0}w$ is large, then $ A\left( \alpha\right) $ is concentrated around $\alpha _{0}$, and if
$\partial_{\alpha}\beta\left( \alpha \right) $ does not vary
too quickly in the vicinity of $\alpha _{0}$, we obtain the
well-know crude approximation:
\begin{equation}
\tan \psi  \sim -\partial_{\alpha}\beta\left(
\alpha _{0}\right)  \label{gracu}
\end{equation}
Of course, the formula (\ref{gracu}) can no longer hold if
$\partial_{\alpha} \beta\left( \alpha \right) $ is not analytic
near $\alpha _{0}$, i.e. when $\alpha _{0}$ is a branch point. We
shall encounter this case in the following section.

In order to give a geometric interpretation of this result, let us remark
that $\left( \partial_{\alpha} \beta \left( \alpha _0 \right) ,-1\right) $ is a
vector that is normal to the dispersion curve at wavelength $\lambda $. So
that we retrieve the well-known fact that for a spatially large beam, the direction of
propagation is given by the normal to the isofrequency Bloch diagram.

{\it We shall see in the following that this result is in general no
longer true in finite 2D structures}.

\subsection{Beam propagation in a 2D photonic crystal}

The crystal is described as a stack of gratings and we assume that
in the spectral domain defined by the above beam, the ratio between
the wavelength and the period $d$ of the gratings is such that
there is only one reflected and one transmitted order for the
grating structure. Then the propagating reflected and transmitted
fields can be expressed as:
\begin{eqnarray}
u_{d}\left( x,y\right) =\int A\left( \alpha \right) r_{N}\left( \alpha
\right) e^{i\left( \alpha x-\beta y\right) }d\alpha \\
u_{t}\left( x,y\right) =\int A\left( \alpha \right) t_{N}\left( \alpha
\right) e^{i\left( \alpha x+\beta y\right) }d\alpha
\end{eqnarray}

Once the reflection and transmission coefficients $\left(
r_{N},t_{N}\right) $ are known (by using a rigorous numerical
method, for instance the Fourier Modal Method (FMM) \cite{FMM})
there exists a unique unimodular real $2\times 2$ matrix
$\mathbf{T}_{N}$ \cite{moi1} with real coefficients satisfying
relation (\ref{transfert}): it is the dressed transfer matrix of the
total structure \cite{moiprl}. We have:

\begin{equation}
\mathbf{T}_{N}=\left(
\begin{array}{cc}
\phi _{11} & \phi _{12} \\
\phi _{21} & \phi _{22}
\end{array}
\right) \left(
\begin{array}{cc}
e^{iNh\widetilde{\beta }_{N}} & 0 \\
0 & e^{-iNh\widetilde{\beta }_{N}}
\end{array}
\right) \left(
\begin{array}{cc}
\phi _{11} & \phi _{12} \\
\phi _{21} & \phi _{22}
\end{array}
\right) ^{-1}
\end{equation}
where the phase $\widetilde{\beta }_{N}$ is the renormalized Bloch phase for the global structure \cite{moiprl}.
From this matrix, the reflection and transmission coefficients can be written in the following form:
\begin{equation}
r_{N}=\frac{\left( e^{2i\widetilde{\beta }_{N}Nh}-1\right) f}
{e^{2i\widetilde{\beta }_{N}Nh}-g^{-1}f}\, ,\,\,\, t_{N}=
\frac{e^{i\widetilde{\beta }_{N}Nh}\left( 1-g^{-1}f\right) }
{e^{2i\widetilde{\beta }_{N}Nh}-g^{-1}f}
\end{equation}

For a stratified medium with homogeneous layers, the reduced
transfert matrix satisfies rigorously the relation:
$\mathbf{T}_{1}^{N}=\mathbf{T}_{N}$ for all $N$. However, for a two
dimensional photonic crystal, this relation tends to become false as
the number of periods is increased: this is due to the fact that
matrix $\mathbf{T}_{1}$ does not take the evanescent waves into
account. Consequently, as the thickness of the device increases the
discrepancy between $\mathbf{T}_{1}^{N}$ and $\mathbf{T}_{N}$
increases as well. This remark has a crucial importance for our
study as, in general, the derivative of the phase $\partial_{\alpha}
\widetilde{\beta }_{N}$ is not equal to $\partial_{\alpha} \beta $.

By definition $tr\left( \mathbf{T}_{N}\right)
=2\cos \left( Nh\widetilde{\beta }_{N}\right) $, so that:
\[
\partial _{\alpha }tr\left( \mathbf{T}_{N}\right) =
-2Nh  \partial_{\alpha} \widetilde{\beta }_{N} \sin \left( Nh\widetilde{\beta }_{N}\right) =
\mp Nh  \partial_{\alpha} \widetilde{\beta }_{N}\sqrt{4-tr\left( \mathbf{T}_{N}\right) ^{2}}
\]
this provides us with a numerical method for computing
$\left|  \partial_{\alpha}\widetilde{\beta }_{N}\right| $, the sign is unambiguously fixed using the
fact that $\left| g\right| <1$ and is associated with the eigenvalue
$e^{iNh\widetilde{\beta }_{N}}$.

We have reduced the problem of computing the transmitted field to
the one-dimensional case, and thus we can write:
\begin{equation}
u_{0}^{t}\left( x,Nh\right) =\int \widetilde{A}\left( \alpha \right)
e^{iNh\widetilde{\beta }_{N}\left( \alpha \right) }e^{i\alpha x}d\alpha
\end{equation}

We can now give the main result of this paper, whose proof is given
in Appendix 3. We assume that the beam is spatially large (i.e.
$k_{0}w\gg 1$). Then two cases may be encountered with respect to
the dispersion curve. For the mean angle of incidence of the beam (corresponding to $\alpha_0$),
the curve is either regular (i.e the slope is not infinite ), or it
is singular, i.e. the tangent to the curve is vertical. The shift of
the beam is then described accordingly:

\begin{enumerate}
\item If the tangent is not vertical, i.e.
$\left|  \partial_{\alpha} \widetilde{\beta }_{N}(\alpha_{0}) \right| <+\infty $
then the angle of refraction $\psi$ of the beam inside the structure is given by:
\begin{equation}
\tan \psi \sim - \partial_{\alpha}\widetilde{\beta }_{N} (\alpha _{0}),
\end{equation}
\item if the slope is infinite, i.e. $\left| \partial_{\alpha} \widetilde{\beta }_{N}(\alpha_{0}) \right| =
+\infty $ then the angle of refraction $\psi$ of the beam
inside the structure is given by:
\begin{equation}
\tan \psi \sim -C_{N}\frac{2^{3/4}}{\sqrt{\pi }}\Gamma \left( \frac{5}{4}\right)
 \sqrt{\sin \theta _{1}+\sin \theta _{0}}\sqrt{k_{0}w},  \label{contra}
\end{equation}
where $\alpha _{1}=k_{0}\sin \theta _{1}$ is the maximum of
$\widetilde{A}\left( \alpha \right) ^{2}\left( \alpha-\alpha
_{0}\right) ^{3/2}$ and $C_{N}$ is a constant such that $\beta_N
\sim C_N \sqrt{\alpha^2-\alpha_0^2}$ near $\alpha_0$.
\end{enumerate}

In the second result, the geometry of the structure
and its electromagnetic parameters enter in the constant $C_{N}$.
For a sufficiently large beam, $\theta _{1}\sim \theta _{0}$.

Two important properties should be noted in that case. The obvious
one is that the shift does not tend to infinity when the normal to
the dispersion curve tends to the horizontal axis, a fact that was
of course expected, but which shows that the normal to the
renormalized dispersion curve gives the direction of the beam, only
if $ \partial_{\alpha}\widetilde{\beta }$ does not vary too quickly in the
vicinity of the mean angle $\alpha _{0}$. Two parameters are in fact
needed for a complete description of the situation: the normalized
waist $k_{0}w$ and the derivative of the phase $ \partial_{\alpha}\widetilde{\beta}\left( \alpha _{0}\right) $. The above result only gives
the asymptotic behavior for the separate parameters.

The second important point is the dependence of the shift with
respect to the normalized waist, a situation which was not
encountered in the first case. In order to understand this point,
one should note that there is here a guided mode in the structure,
i.e. a pseudoperiodic mode whose wavevector has a null vertical
component. Of course, for a finite size structure, the uniqueness of
the scattering problem implies that guided modes are associated with
complex values of $\alpha $ which are poles of the transmission
coefficient. The finiteness of the structure provokes a splitting of
the eigenvalue $\alpha _{0}$ into two complex values \cite{moi2}
corresponding to a zero and a pole of the reflexion coefficient.
When such a structure is illuminated by a plane wave under the
incidence $\alpha _{0}$ the transmission shows a Fano profile
indicating the excitation of the lossy mode. When the incident light
is a beam, the behavior of the field resembles that of a plane wave
in the limit $k_{0}w\gg 1$, therefore the displacement of the
barycenter towards infinity is associated with a spreading of the
transmitted beam and thus, precisely because of the spreading, the
very notion of barycenter of the transmitted beam loses its physical
meaning.

\section{Numerical examples}

In the following, we present some numerical computations
illustrating the various situations described by the above results.
We will denote by:
\begin{itemize}
\item $\Delta$ the shift computed by direct numerical computations of the
fields.

\item $\Delta_{\beta}$ the shift computed through the isofrequency
 dispersion diagram.

\item $\Delta \widetilde{_{\beta }}$ the shift computed by use of the effective theory developed in the
previous section.
\end{itemize}

\subsection{Case of a stratified one dimensional medium}

In this subsection, we check the numerical method that allows to
compute the derivative of the Bloch vector of the equivalent
$\mathbf{T}$ matrix and also the formula that gives the shift of the
beam. The structure that we use is just a Bragg Mirror with two
alternating slabs (thicknesses $h_{1}$ and $h_{2}$) in each period.
The $s$ polarized incident monochromatic beam is characterized by
its waist $w=15 \lambda$ and its mean angle of incidence $\theta
_{0}=50^o$. The wavelength is such that $\lambda /h_{1}=2.27$ with
$h_{1}/h_{2}=2$ and the following permittivities for the slabs:
$\varepsilon _{1}=2.1,\varepsilon_{2}=6.25$. In fig.2, we give the
amplitude of the incident, transmitted and reflected fields on the
upper and lower interfaces of the device for $N=15$ periods. The shift of the first
transmitted beam obtained by direct numerical computation is $\Delta
/h_{1}=12.51$ whereas the shift obtained by computing the Bloch
coefficient is $\Delta _{\beta}/h_{1}=12.52$ and finally, the
numerical computation described in the above section gives $\Delta
\widetilde{_{\beta }}/h_{1}=12.52$ hence, as expected, a perfect
agreement with the Bloch approach. In order to complete this
verification, we now use a $p$-polarized incident beam with
$w=15 \lambda,\lambda /h_{1}=4.75,\theta _{0}=47.5^o$ and the parameters
$\varepsilon _{1}=10.89,\varepsilon _{2}=1,h_{1}/h_{2}=1 $. Fig.3,
shows the amplitude of the transmitted and reflected fields. This
time, we obtain $\Delta /h_{1}=62.7$, $\Delta_{\beta
}/h_{1}=62.7209,$ $\Delta \widetilde{_{\beta }}/h_{1}=62.7209$,
which confirms the validity of our approach for that straightforward
situation.

\subsection{Stack of gratings}

We are now in a position to apply our theoretical approach to the
more complex situation of a stack of gratings. We recall that we
assume that the wavelength is such that there is only one Bloch mode
inside the structure and one transmitted order and one reflected
order. For all the following numerical experiments the field is s-polarized.

The photonic crystal is a stack of $7$ lamellar gratings with
inverted contrast: $\varepsilon _{1}=\varepsilon _{ext}=11.56$ and
$\varepsilon_{2}=1$( $d_{1}/d=h_1/d=1/2$ and $h=d$, see fig.2 for
notations). We compute the field for an incident beam
($w=12.5 \lambda,\lambda /d=2.2,\theta =50^o$) using the Fourier Modal
Method \cite{FMM}. The amplitudes of the field on the upper and
lower interfaces are given in fig.4. The Bloch diagram is given in
fig.5. The shift of the transmitted beam obtained directly from this
computation (through the envelope) is $\Delta /d\sim 11.25$ the
shift computed from the isofrequency dispersion diagram is $\Delta
_{\beta }/d=2.45$ and the shift obtained from the effective theory
is $\Delta _{\widetilde{\beta }}/d=11.4$. Therefore, we see that we
have an error factor of $4.5$ by neglecting the evanescent waves.

The effective theory also applies when contra-propagative Bloch
modes exist in the structure, these modes authorizing super-prism
phenomena. As an example, we consider a 2D PhC made of $5$ lamellar
grating layers ($\varepsilon_{1}=9,\e_2=\e_{ext}=1,d_{1}/d=0.77,h=d,h_{1}/d=1/4$).
The isofrequency Bloch
diagram of the structure is given in fig.6 for $\lambda /d=2$.
There is a zone of contra-propagating Bloch modes around
$\alpha_{0}=1.6$ ($\theta \sim 40^o$). The structure is illuminated by a
monochromatic gaussian beam ($w=10 \lambda,\alpha _{0}=1.6,\lambda /d=2$).
 The map of the field is given in fig.7,
where it is seen that the shift of the transmitted beam is negative,
the amplitude of the field on the upper and lower faces are given in
fig.8. The shift obtained by the direct numerical computation is
$\Delta \sim -9.3$, the shift obtained from the Bloch diagram is
$\Delta _{\beta }/d=-3.3$ whereas $\Delta _{\widetilde{\beta
}}/d=-9.9$. Once more, we find an excellent agreement between the
direct computation and the effective theory, whereas the predictions
of Bloch theory are quite false.

Let us turn now toward a structure in which a guided
contra-propagative mode do exist, this corresponds to the situation
$2$ in the proposition of the preceding section. In other words,
there exists a mode with an horizontal wave vector. The parameters
are the following: $\varepsilon_{1}=7.84,d_1/d=0.4,h_{1}/d=0.7,h_{2}/d=0.3,\lambda /d=2.1,w=50\lambda$, and
there are $7$ layers in the PhC. For this structure the
contra-propagative mode is obtained for $\theta _{0}=39^o$. Here,
the parameter $C_{7}$ which represents the behavior of $\beta \left(
\alpha \right) \sim C_{7}\sqrt{\alpha ^{2}-\alpha _{0}^{2}}$ is
obtained by a fitting of the results of the direct numerical
computation, we obtained $C_{7}=0.164$. We have plotted in fig. 9
$\left( d/\Delta \right) $ versus $d/\left( \alpha -\alpha
_{0}\right) $ where it is clearly seen that the shift converges
towards a limit value. Using formula (\ref {contra}), we obtain
$\Delta _{\widetilde{\beta }}/d\sim -26$ in fair agreement with the
numerical shift $\Delta /d\sim -27$.

One should not think, however, that the non-renormalized Bloch
diagram, i.e. that of the infinite periodic structure, cannot
provide us with accurate results. It suffices to think of the
homogenization regime, where the stack of gratings behaves as a
stratified medium. For instance, we use a stack of $7$ gratings
($h_{1}/d=1/2,h=d,\varepsilon =2.1$) and a beam with
parameters: $w=25\lambda,\lambda /d=4,\theta _{0}=40^o$.
 The field amplitudes on the upper and lower faces are given in fig. 10,
where it can be seen that the oscillations are quite limited showing that we
are indeed in the homogenization regime. The shift of the beam is
$\Delta/d=8$, and we have $\Delta _{\beta }/d=7.985$ and
$\Delta _{\widetilde{\beta}}/d=7.985$. In that case, both predictions agree. This situation is due to
the fact that the field inside the PhC can be represented by Bloch modes
only. This situation may happen outside the homogenization regime.

\section{Conclusion}

We have developped an effective medium approach to describe beam
propagation inside a photonic crystal. This effective theory takes
into account the evanescent waves, which are completely skipped if
one uses only Bloch waves to describe wave propagation in the
crystal. The importance of these evanescent waves are put into light
by the computation of the shift of the transmitted beam. We show for
some examples that the predictions obtained by using only the
dispersion diagram may be false. These results
emphasize the difference between the band theory for the
Schr\"{o}dinger equation (i.e. the propagation of electrons in
periodic potentials), where the boundary of the crystal is
irrelevant, and the scattering of electromagnetic waves by photonic
crystals where boundary effects are of crucial importance. 
We have developed elsewhere  \cite{physicae}  some theoretical tools, that should hopefully permit
to obtain a clearer insight into the role of evanescent waves. Work
is in progress in that direction.

\newpage

\begin{center}
{\LARGE Appendix 1}
\end{center}

\textbf{Lemma}

\textit{Let }$f$\textit{\ be a real even function continuously
differentiable near }$ \alpha _{0}$ {\it and such that} $f(\alpha_0)=0$.
\textit{Then if }$f^{\prime }$\textit{\ is square integrable near }$\alpha_{0}$
\textit{there holds}
\begin{equation}
\frac{f\left( \alpha \right) }{\sqrt{\alpha ^{2}-\alpha _{0}^{2}}}=O\left(
1\right)
\end{equation}

Proof:

Let us write: $f\left( \alpha \right) =f\left( \alpha ^{\prime }\right)
+\int_{\alpha ^{\prime }}^{\alpha }f^{\prime }\left( t\right) dt$. Then
$\left( f\left( \alpha \right) -f\left( \alpha ^{\prime }\right) \right)
^{2}\leq \left( \int_{\alpha ^{\prime }}^{\alpha }f^{\prime }\left( t\right)
dt\right) ^{2}$ and then by Cauchy-Schwarz inequality: $\left( f\left(
\alpha \right) -f\left( \alpha ^{\prime }\right) \right) \leq \sqrt{\alpha
-\alpha ^{\prime }}\sqrt{\left( \int_{\alpha ^{\prime }}^{\alpha }\left(
f^{\prime }\left( t\right) \right) ^{2}dt\right) }$ and the proposition
follows.

\newpage

\begin{center}
{\LARGE Appendix 2}
\end{center}

Applying the same reasoning as for the stratified medium, we get:
\[
G_{t}=-Nh\int \widetilde{A}^{2}\left( \alpha \right)
 \partial_{\alpha}\widetilde{\beta }_{N}\left( \alpha \right) d\alpha /E_{t}
\]

Assuming that $u_{0}^{t}$ has a first moment, we obtain, by Parseval
equality:

\begin{eqnarray*}
\int x\left| u_{0}\left( x\right) \right| ^{2}dx &=&\frac{1}{i}\int
\widetilde{A}\left( \alpha \right) \widetilde{A}\left( \alpha \right)
^{\prime }d\alpha +\int \widetilde{A}^{2}\left( \alpha \right)
 \partial_{\alpha}\widetilde{\beta }_{N}\left( \alpha \right) d\alpha \\
&=&\int \widetilde{A}^{2}\left( \alpha \right)
 \partial_{\alpha}\widetilde{\beta }_{N}\left( \alpha \right) d\alpha
\end{eqnarray*}
and the angular shift is given by (cf. fig. 1):
\[
\tan \psi =-\frac{\int \widetilde{A}^{2}\left( \alpha \right)
 \partial_{\alpha}\widetilde{\beta }_{N}\left( \alpha \right) d\alpha }
{\int \widetilde{A}^{2}\left( \alpha \right) d\alpha }
\]

The dispersion diagram is described locally by $\beta =\phi \left( \alpha
\right) $ and, in the vicinity of the branch point $\alpha _{0}$, we can
write from the lemma proved in Appendix 2:
\[
\phi \left( \alpha \right) =C\sqrt{\alpha ^{2}-\alpha _{0}^{2}},\alpha \geq
\alpha _{0}
\]
the shift is then given by:

\begin{equation}
\tan \psi =-C\frac{\int \widetilde{A}^{2}\left( \alpha \right) \frac{\alpha
}{\sqrt{\alpha ^{2}-\alpha _{0}^{2}}}d\alpha }{\int \widetilde{A}^{2}\left(
\alpha \right) d\alpha }=-C\sqrt{\sin \theta _{1}+\sin \theta _{0}}
\frac{2^{3/4}}{\sqrt{\pi }}\Gamma \left( \frac{5}{4}\right) \sqrt{kW}
\end{equation}
where $\alpha _{1}=k\sin \theta _{1}$ is a maximum of $\alpha \rightarrow
\widetilde{A}\left( \alpha \right) ^{2}\left( \alpha -\alpha _{0}\right)
^{3/2}$.

\newpage
\begin{figure}[h]
    \includegraphics*[width=6cm,height=6cm]{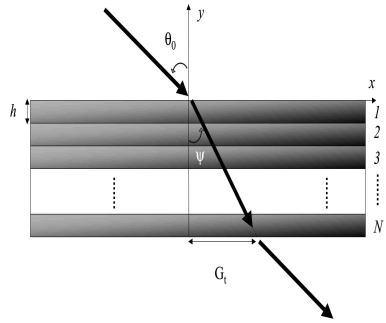}
    \caption{sketch of the photonic crystal}
    \end{figure}
\begin{figure}[h]
    \includegraphics*[width=6cm,height=6cm]{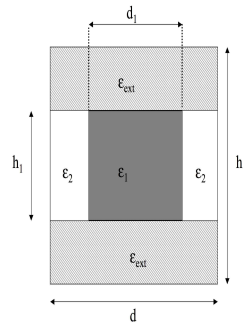}
    \caption{Basic cell of the photonic crystal used in the numerical experiments.}
     \end{figure}
\begin{figure}[h]
    \includegraphics*[width=6cm,height=6cm]{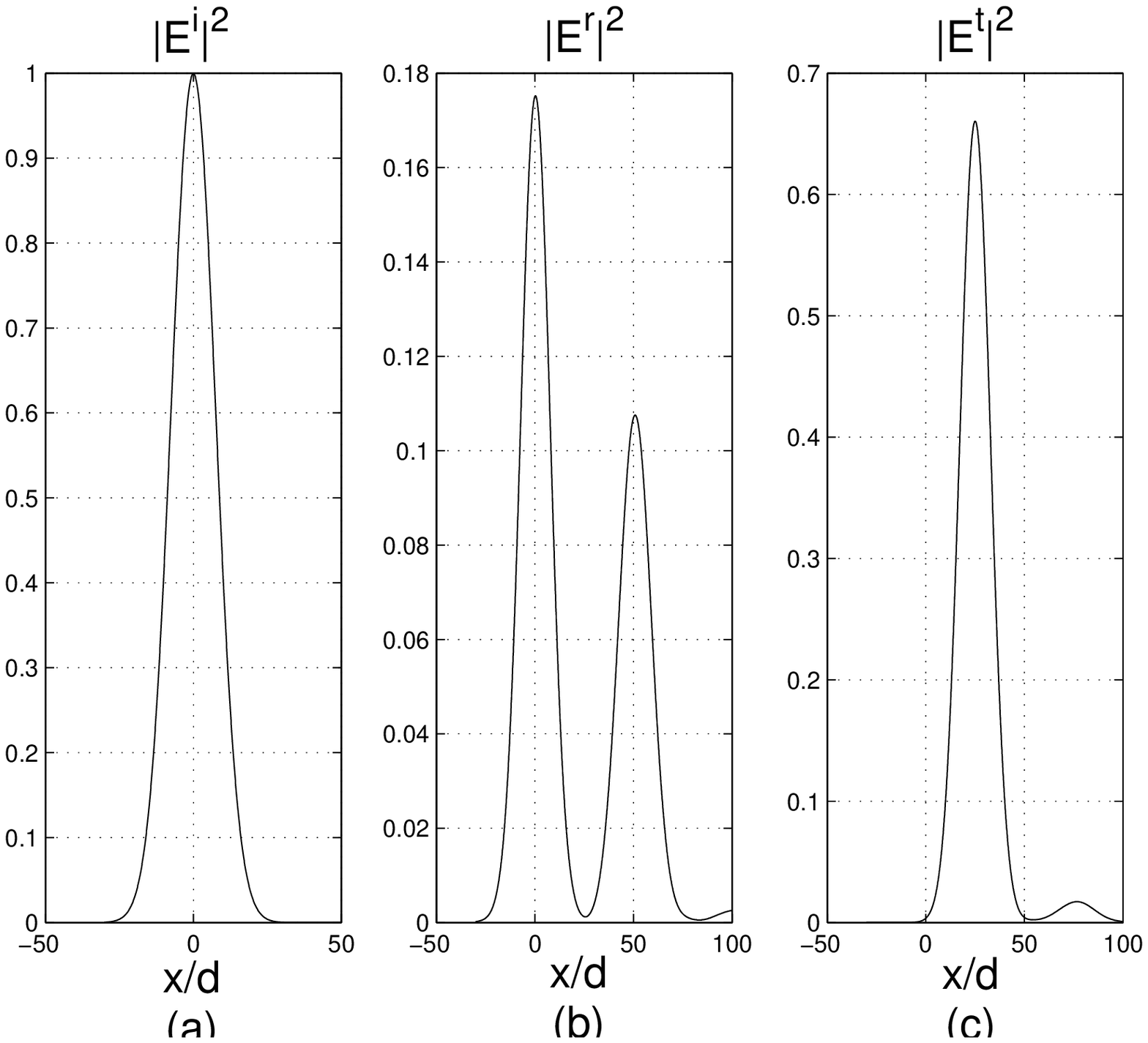}
    \caption{Incident (a) transmitted (b) and reflected (c) field
intensities for an s-polarized incident field on a 1D structure.}
   \end{figure}
\begin{figure}[h]
    \includegraphics*[width=6cm,height=6cm]{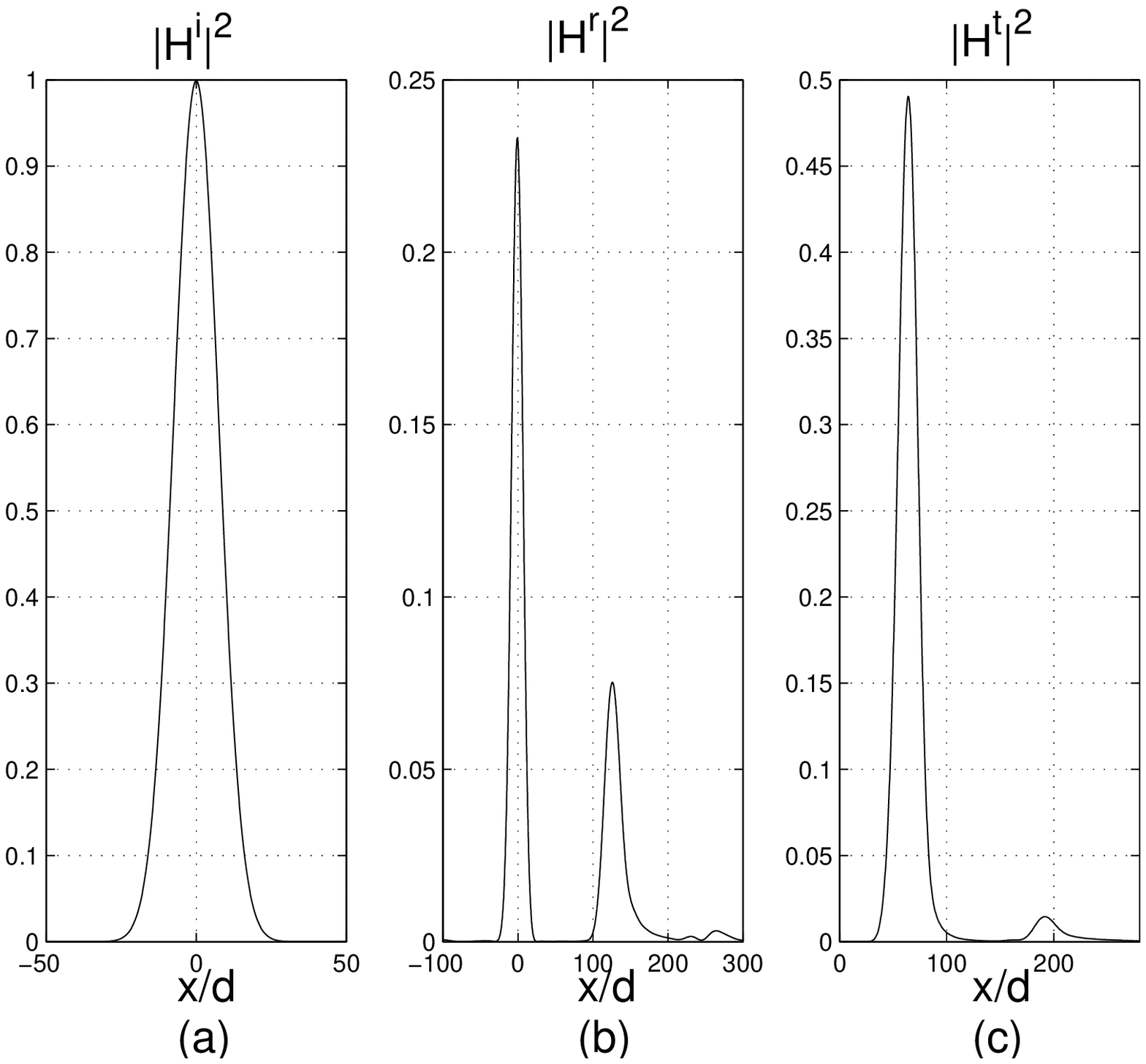}
    \caption{Incident (a) transmitted (b) and reflected (c) field
intensities for a p-polarized incident field on a 1D structure}
 \end{figure}
\begin{figure}[h]
    \includegraphics*[width=6cm,height=6cm]{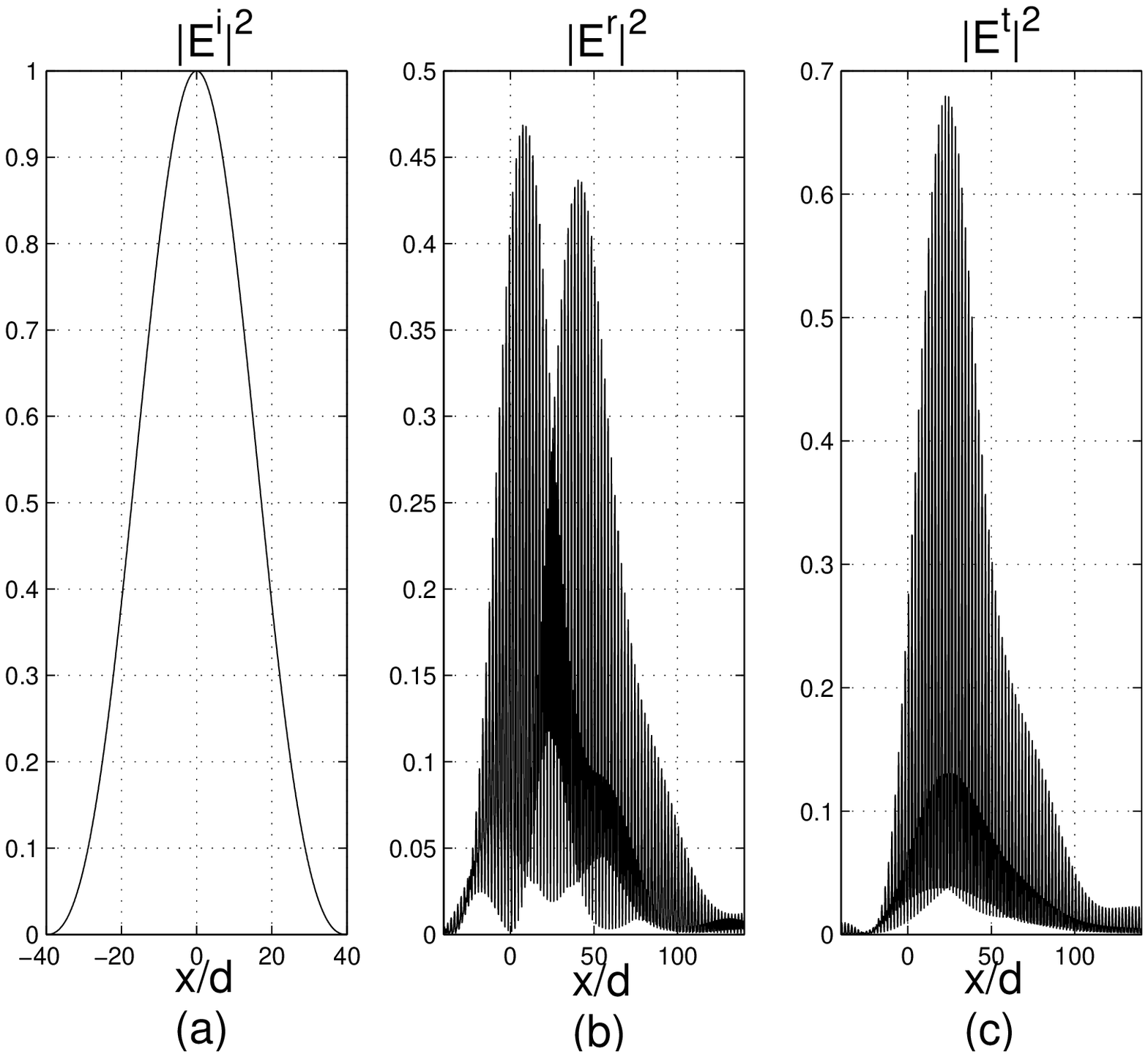}
    \caption{Incident (a) transmitted (b) and reflected (c) field
intensities for an s-polarized beam on a 2D PhC.}
 \end{figure}
\begin{figure}[h]
    \includegraphics*[width=6cm,height=6cm]{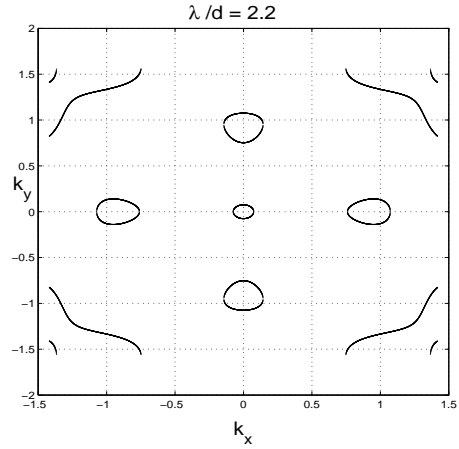}
    \caption{Bloch diagram for the inverted contrast photonic crystal.}
 \end{figure}
\begin{figure}[h]
    \includegraphics*[width=6cm,height=6cm]{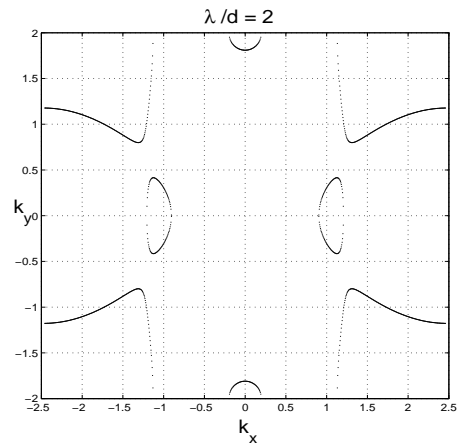}
    \caption{Bloch diagram for the photonic crystal allowing contra-propagating modes.}
 \end{figure}
\begin{figure}[h]
    \includegraphics*[width=6cm,height=6cm]{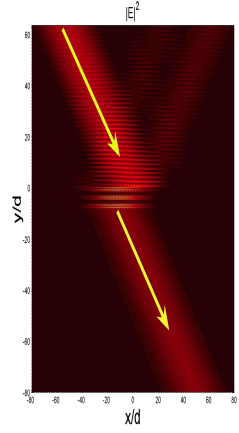}
    \caption{Map of the electric field for the contra-propagating mode.}
 \end{figure}
\begin{figure}[h]
    \includegraphics*[width=6cm,height=6cm]{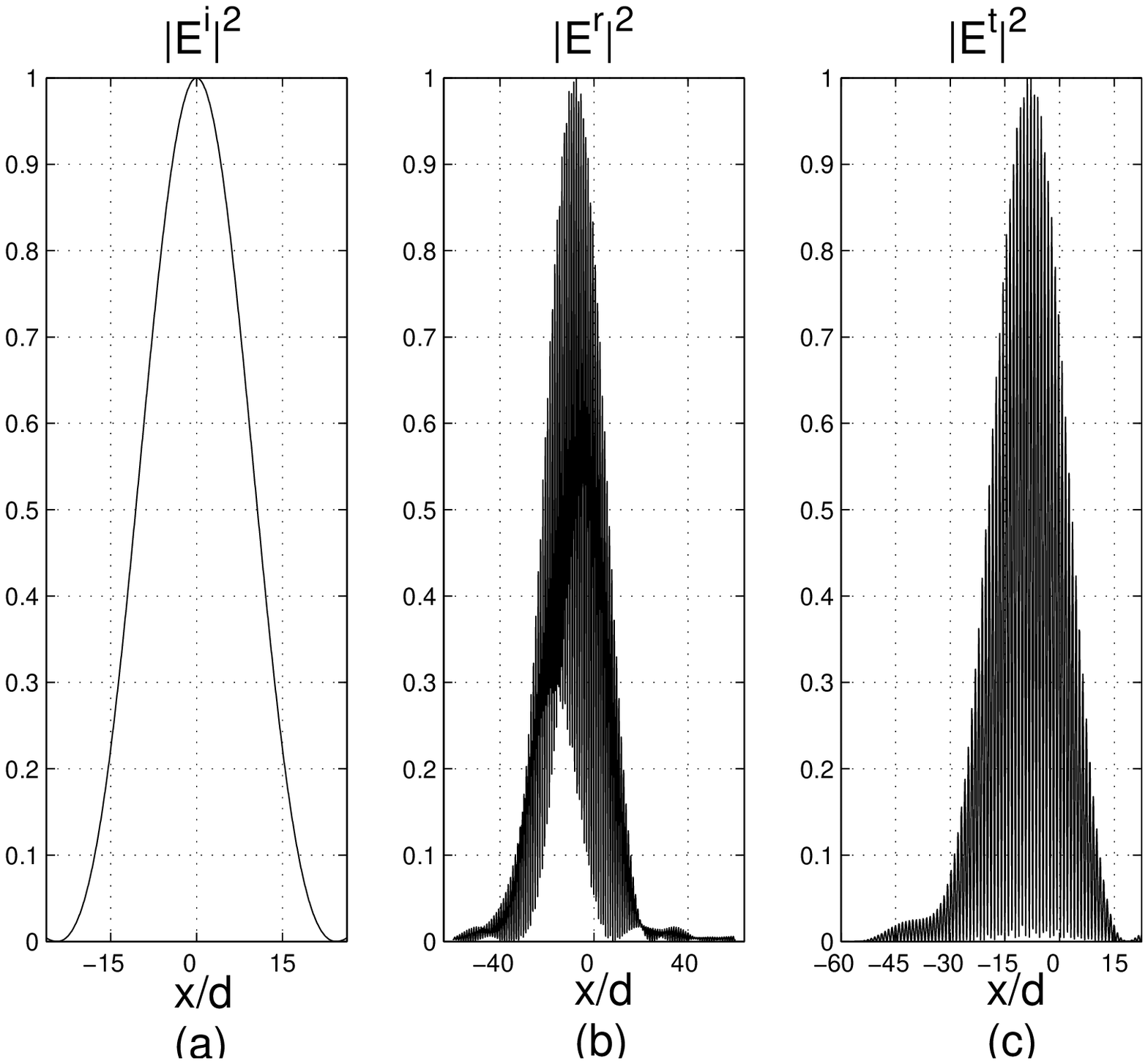}
    \caption{Incident (a) transmitted (b) and reflected (c) field
intensities for an s-polarized beam on a 2D PhC.}
 \end{figure}
\begin{figure}[h]
    \includegraphics*[width=6cm,height=6cm]{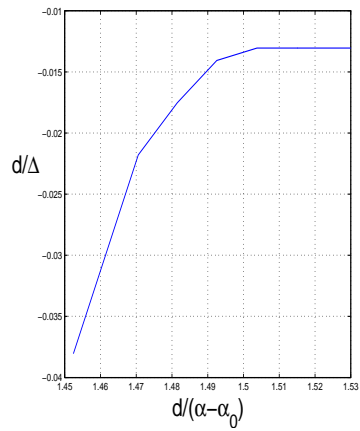}
    \caption{Evolution of the shift with respect to the Fourier
variable.}
    \end{figure}
\begin{figure}[h]
    \includegraphics*[width=6cm,height=6cm]{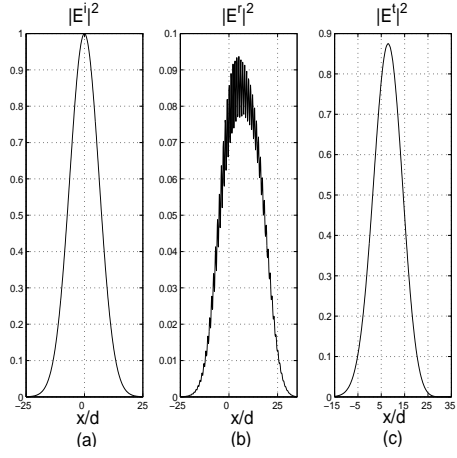}
    \caption{Incident (a) transmitted (b) and reflected (c) field
intensities in the homogenization regime.}
 \end{figure}

\end{document}